\documentclass[draft,showpacs,preprintnumbers,amsmath,amssymb]{revtex4}
\usepackage{graphicx}% Include figure files
\usepackage{dcolumn}% Align table columns on decimal point
\usepackage{bm}% bold math

\begin{document}

\title{o-Positronium scattering off H and He} 

\author{Simone Chiesa}\email{Simone.Chiesa@unimi.it}
\author{Massimo Mella}\email{Massimo.Mella@unimi.it} 
\affiliation{
Dipartimento di Chimica Fisica ed Elettrochimica,Universita' degli Studi
di Milano, via Golgi 19, 20133 Milano, Italy\\}
\author{Gabriele Morosi}\email{Gabriele.Morosi@uninsubria.it}
\affiliation{Dipartimento di Scienze Chimiche, Fisiche e Matematiche,
Universita' dell'Insubria,
via Lucini 3, 22100 Como, Italy}

\begin{abstract}
Exploiting an approach similar to the R-matrix theory, the diffusion 
Monte Carlo method is employed to compute phase shifts and threshold 
cross sections for the elastic scattering of o-positronium off light atoms. 
Results are obtained for Ps-H and Ps-He as representative cases of open and
closed shell targets.
The method allows for an exact treatment of both correlation and exchange 
interactions, and represents
the most promising approach to deal with these effects in more complicated 
targets. In particular the Ps-He threshold cross section, computed in a many 
body framework for the first time, represents a standard by which past and 
future numerical and experimental estimates can be judged.
%In this respect, it is worth 
%stressing how our value shows the inaccuracy 
%of all the most recently proposed and disagreeing estimates for the same quantity. 
\end{abstract}

\maketitle 

Positronium (Ps) scattering off atomic and molecular        
targets has an overwhelming importance if an understanding
of the interaction mechanism between an overthermal Ps and a 
condensed matter environment is required \cite{moge}. For instance, by means
of elastic and inelastic cross sections, it may be possible to model
energy transfers from Ps to the surroundings or to describe
the trapping of Ps in a free volume cavity.
Despite its long history \cite{Drach1,Drach2,Page,Drach3,Fraser,Canter}, 
and even in the case of light atoms,
some quantitative aspects of the process still
remain controversial and have recently been addressed by a number of 
authors, both experimentally \cite{Colema,Nagash,Garmer,Skalsey} 
and theoretically \cite{Campb,Adhik,Biswas,Blackw,Jim}.  
From the computational point of view, the difficulties 
with which almost every method is faced are related to the
composite nature of both target and projectile. As a consequence,
sensible results can be obtained only if  
correlation and exchange effects are properly treated. 
Moreover, an accurate description of the correlation effects between the 
target electrons and the positron in the Ps is important
in computing "pick off" annihilation rates.
These effects have only recently been computed consistently
for the case of positronium scattering off hydrogen and
positronium atoms \cite{Jim,Shumw}. However, the full {\it ab initio} treatment 
(i.e. without the use of exchange or correlation model potentials) of 
systems with more than two electrons still represents a 
formidable task. A glance to the recent literature on
bound systems containing a positron reveals a practically
identical situation with only small numbers of electrons
treated explicitly.  
In this context, it has been shown by several authors \cite{jiang,maxcor} that
flexible and accurate computational techniques  
for small and medium size systems are provided by the family of quantum Monte
Carlo (QMC) methods.
%The diffusion
%Monte Carlo (DMC) method we employed in this work directly samples a set of point in configurational
%space (also known as walkers) whose distribution is proportional to the 
%exact ground state wave function. This distribution allows one to compute
%exactly the total energy of the system and also the expectation values
%of many other observable. Since the DMC simulation method is based on
%the formal equivalence between the Schroedinger equation in imaginary time
%and a diffusion-reaction equation, the sampled wave function must be definite
%positive. When fermionic systems are concerned, one is forced to divide
%the configurational space into equivalent non connected volumes in order
%to mimic the presence of nodal regions having opposite sign. This task
%is usually accomplished exploiting the nodal surface of a suitable optimized
%wave function to confine the walker displacement. If the nodal surface are
%exact, i.e. are the same of the exact wave function, then all the expectation
%values are also exact. Conversely, if the nodal surfaces are just an
%approximation of the correct ones, the computed mean energy 
%$\langle E \rangle _{DMC}$ is an upper 
%bound to the ground state eigenvalue $E_0$. The difference between $E_0$ and
%the $\langle E \rangle _{DMC}$ is usually referred to as ``nodal error''
%and its relative value commonly spans the $[10^{-5},10^{-4}]$ a.u. range
%\cite{arnerev}.
Among them, the diffusion Monte Carlo (DMC) scheme represents
the most powerful approach to study strongly correlated systems
thanks to its ability
to sample a distribution proportional 
to the exact ground state wave function of a given Hamiltonian. 
Where fermions are concerned, the
antisymmetric nature of the wave function and its 
consequent non-positiveness imply the introduction of 
a bias known as nodal error. For a given energy $\epsilon$, 
the nodal error $\Delta \epsilon$, which disappears if
the nodal surfaces of the exact wave function are known, has a value which 
commonly spans the range $\Delta \epsilon / \epsilon \in [10^{-5},10^{-4}]$ 
\cite{arnerev}. Unless otherwise specified, the
following results have been computed in the DMC framework.
%Since the QMC methods were devised to deal with bound ground states, 
%their application to collisional problems seems apparently precluded by 
%the fact that the scattering stationary wave function is not 
%normalizable and thus not compatible with a classical distribution.
 
The application of QMC methods to scattering problems was 
independently proposed in the eighties in two 
pioneering papers by two groups in the field of 
nuclear physics \cite{koonin,Carlson}. Their ideas have been recently 
applied to the exciton-exciton scattering problem \cite{Shumw},
thus providing the first accurate
calculation for the Ps-Ps system.
This approach, which closely resembles the original idea behind 
the R-matrix theory of Wigner \cite{wigner}, is 
briefly summarized in what follows for the case of an elastic collision. 
%The crucial point is to divide the space around the target in
%two regions where the problem can be solved exactly
%(either analytically or computationally), and then match
%the two solutions requiring the wave function to have the
%necessary continuity properties at the boundary. 
We define $\mathbf{r}_{AB}=\mathbf{R}_{A}-\mathbf{R}_{B}$ as the
relative position of the centers of mass of $A$ and $B$ and $p$ and $\mu$ as
their relative momentum and reduced mass. The choice of a boundary 
$r_{AB}=\cal{R}$ under the condition $V_{int}({\cal R}) \ll\frac{p^{2}}{2\mu}$,
allows one to approximate the wave function in the region $r_{AB} > \cal{R}$ 
with the asymptotically exact form
\begin{equation}
\Psi= \mathcal{A}\left[\Psi_{A}(\mathbf{s}_{A})\Psi_{B}(\mathbf{s}_{B})\frac{\Phi_l(r_{AB})}{r_{AB}} Y_{lm}\right]
\label{factform}
\end{equation}
\noindent
$\mathcal{A}$ is the antisymmetrization operator, $\mathbf{s}_{A}$ and 
$\mathbf{s}_{B}$ the internal coordinates of the two
separate fragments, $\Psi_{A}$ and $\Psi_{B}$ their wave functions and 
$Y_{lm}$ and $\Phi_l$ the angular and radial functions 
describing the dynamics of the relative
motion of the two centers of mass.
%($l$ and $m$ are the usual angular momentum quantum numbers). 

The stationary form of $\Phi_l$ can be expressed as  
%\begin{equation*}
%\tilde{\Phi}(\mathbf{r}_{AB})=\frac{\Phi(\mathbf{r}_{AB})}{r_{AB}} 
%\end{equation*}
\begin{equation}
\Phi_l(\mathbf{r}_{AB})=\mathcal{I}_{l}(pr_{AB})+S_{l}(p)\mathcal{O}_{l}
(pr_{AB}) 
\label{relfun}
\end{equation}
%being $l$ and $m$ the usual angular momentum quantum numbers,
where
$\mathcal{I}_{l}$ and $\mathcal{O}_{l}$ are 
Hankel functions, and $S_l(p)$ is the scattering matrix. 
Here, $p$ is connected to 
the total energy by $E_{tot}=\frac{p^{2}}{2\mu}+E_{A}+E_{B}$
where the computation of $E_{A}$ and $E_{B}$, both being bound state energies, 
can be exactly performed employing the DMC method. 
%The $r_{AB} < {\cal R}$ region,
%can be now attacked by a QMC technique imposing the boundary
%condition $\frac{\Phi'({\cal R})}{\Phi({\cal R})}=\mathcal{B}$.
The condition 
$\frac{\Phi'({\cal R})}{\Phi({\cal R})}=\mathcal{B}$
and a {\it corresponding} value of $p$ (i.e. $E_{tot}$) are enough
to  precisely
state the value of $S_l(p)$ in Eq.(\ref{relfun}). 
The link between $\cal{B}$ and $p$ is contained in the
dynamics of the interior region and therefore the computation of $E_{tot}$
in this region, under the same boundary condition on
$r_{AB}={\cal R}$ (i.e. imposing the continuity of the logarithmic
derivative), provides the value of $S_l$. 
Eventually the scattering 
matrix assumes the form
\begin{equation}
S_{l}(p)=-\frac{\mathcal{I}_{l}'(p\mathcal{R})-\mathcal{B}\mathcal{I}_{l}
(p\mathcal{R})}{\mathcal{O}_{l}'(p\mathcal{R})-\mathcal{B}\mathcal{O}_{l}
(p\mathcal{R})} 
\label{scatmat}
\end{equation}
\noindent
Hence, as long as the boundary condition is exactly controlled,
all the relevant information is contained in the interior region.
In this subspace the wave function is normalizable and the energy,
now parametrically dependent on $\cal{R}$, can be studied
by means of one of the QMC techniques e.g. VMC, DMC or one of their variants.
In order to easily fulfill the above  
boundary condition in the DMC framework,
one can choose the value of $\mathcal{B}$ to be 
infinite. From a physical point of view,
this choice corresponds to add a rigid wall located at $\mathcal{R}$
along the distance between the target and projectile centers of mass.
%, whose value
%can be chosen compatibly with the requirement 
%$V_{int}(\mathcal{R} \ll \frac{p^{2}}{2\mu}$.  
%In this approach 
%$E_{tot}$ becomes parametrically dependent on $\mathcal{R}$ and can
%be computed by means of the Diffusion Monte Carlo method.
%Then, once $E_{tot}$ and $\frac{p^{2}}{2\mu}$ are known for a given 
%$\mathcal{R}$, taking the $\mathcal{B}\rightarrow \infty$ limit of 
%Eq. \ref{scatmat} will give the value of $S_{l}$, and hence all the %information
%on the scattering process.
Results will be presented in terms of phase shift $\delta_l(p)$ which
is defined by 
$S_l=e^{2i\delta_l}$ and can be expressed, starting from Eq.(\ref{scatmat})
and taking the $\cal{B}\rightarrow \infty$ limit, as
\begin{equation}
\tan{\delta_l(p)}=\frac{j_l(p\cal{R})}{n_l(p\cal{R})}
\end{equation}
where $j_l$ and $n_l$ are respectively the spherical Bessel and Neumann
functions. 

Before going on we feel that it is worth stressing two important points.
First, one has to satisfy the condition 
$V \ll\frac{p^{2}}{2\mu}$, so an upper limit to the 
sphere radius $\mathcal{R}$ does not exist, 
while it cannot be chosen smaller than some unfortunately not well specified
threshold value. This imposes an upper
limit to the relative kinetic energy. For Ps scattering off 
neutral atoms, the interaction potential between the target and the projectile
dies off as $1/r_{AB}^6$. This short range potential
allows the use of fairly small radii, a possibility not necessarily
available for different colliding fragments.
Secondly, since DMC samples the lowest energy state, it cannot be applied to
scattering problems in presence of a bound state.

%is basically a ground state technique, its straightforward
%application to scattering problems where one or more bound states are present
%is not possible.  If this were the case the method would sample the wave 
%function of the ground state which does not
%carry any information about a possible scattering process.

Both these two points highlight the importance that studying excited 
states could have within this approach. The possibility of raising the energy 
whilst keeping the surface constrain fixed can, in principle, allow the study of every system at any energy.
\begin{figure}[b]
\rotatebox{-90}{\resizebox{!}{0.45\textwidth}{\includegraphics{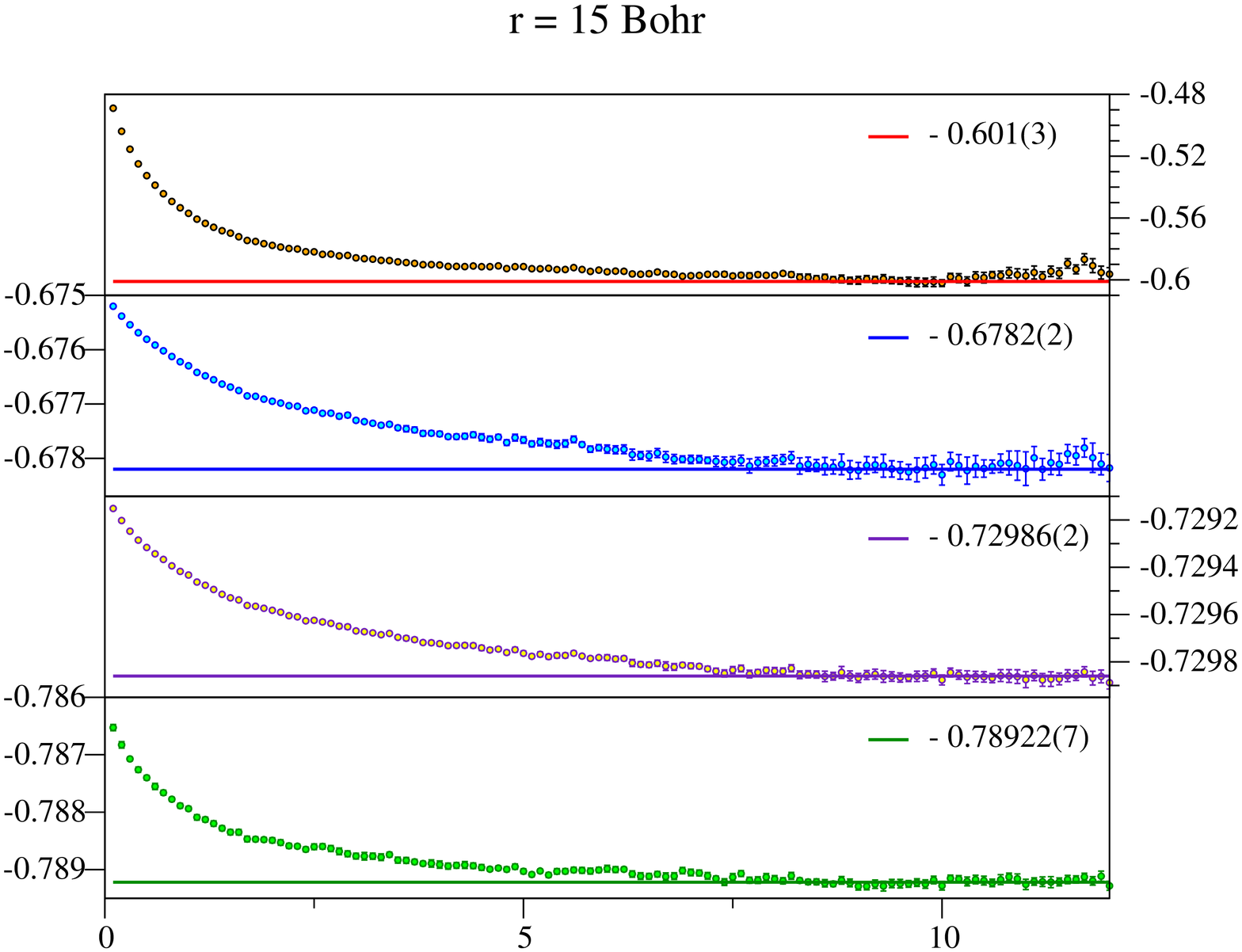}} }
\caption{Energy decay for the first four states of Ps-H system with S=0 
and $\cal{R}$=15 bohr. Note that the energy of the ground state is correctly 
less than -0.75 hartree and coincident with that of the bound state of PsH,
-0.78919 hartree.}
\label{convgraph}
\end{figure}

\begin{figure}
\rotatebox{-90}{\resizebox{!}{0.45\textwidth}{\includegraphics{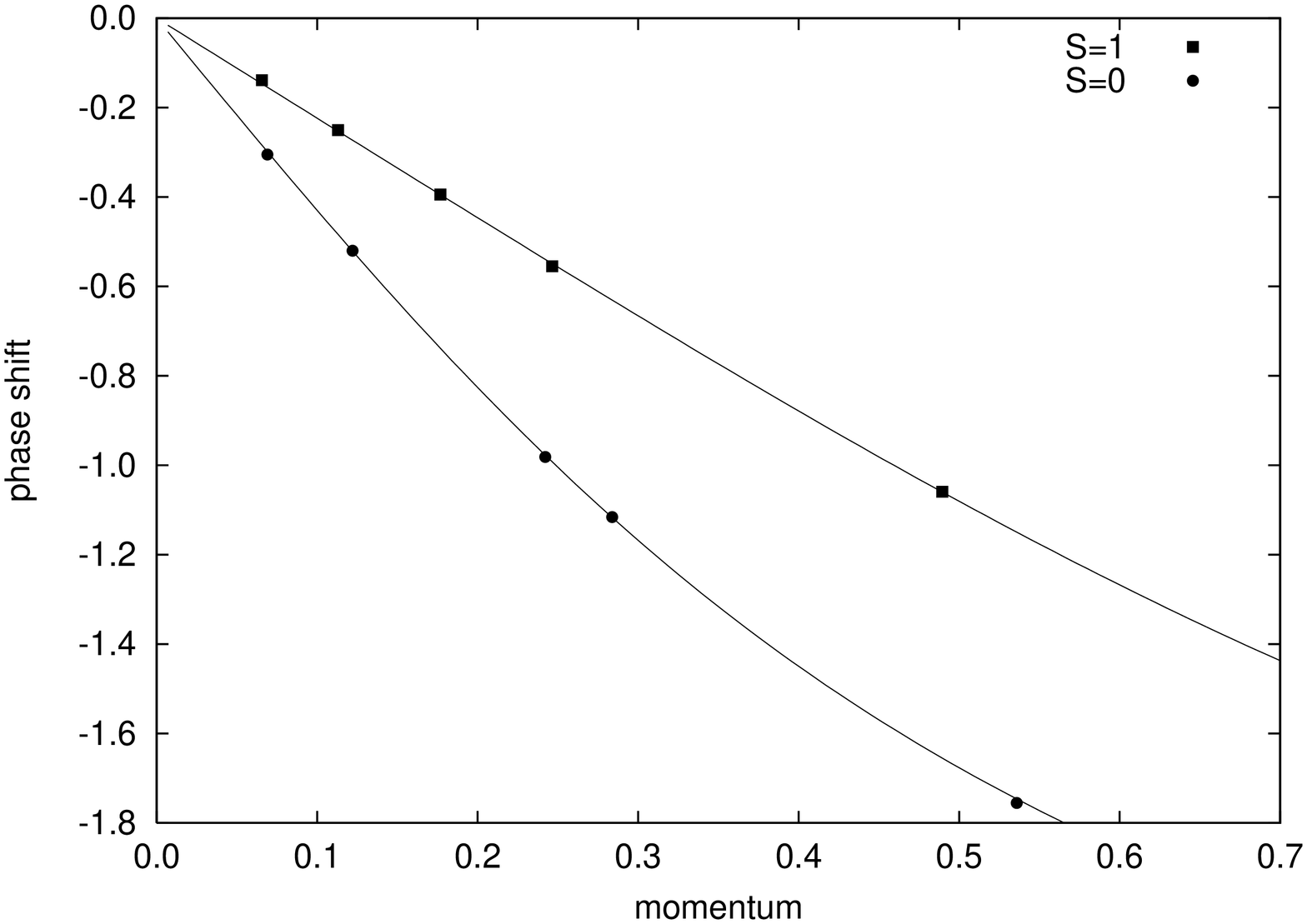}} }
\caption{Phase shift for Ps-H S-wave elastic scattering with total electron 
spin S=1 and S=0. Momentum in atomic units and phase shift in radians.}
\label{HPsgraph}
\end{figure}

In this work, we applied the presented technique to the S-wave scattering 
of positronium off hydrogen and helium dynamically described by the full Hamiltonian
\begin{equation}
H=-\frac{1}{2}\sum_{i=1}^{N_{e}}\nabla^{2}_{i}-\frac{1}{2}\nabla^{2}_{p}-\sum_{i=1}^{N_e}\frac{Z}{r_i}+\frac{Z}{r_p}+\sum_{i>j}\frac{1}{r_{ij}}-\sum_{i}^{N_{e}}\frac{1}{r_{ip}} 
\end{equation}
where $i$ and $j$ refer to electrons, $p$ to the positron and $Z$ to 
the nuclear charge of the atom.  
The spatial part of the scattering wave function has 
been chosen to be of the form
\begin{equation}
\Psi=\mathcal{O}[\Psi_{A}(s_{A})\Psi_{Ps}(r_{1p})\frac{\Phi(r_{PsA})}{r_{PsA}}\phi_{J}(s_{I})]
\end{equation}
where $\Psi_{A}$, $\Psi_{Ps}$, and $\Phi$ have the same meaning as in Eq. 
(\ref{factform}). $\phi_{J}$ is a Jastrow factor for all the pairs 
of particles belonging to different fragments, $s_{I}$ is the set of 
distances for these pairs and $\mathcal{O}$ is the appropriate symmetry 
operator built according to Young diagrams.
In the Ps-H case, the exact internal wave 
function of both fragments is known and $\mathcal{O}$ has the form
\begin{equation}
\mathcal{O}= 1 + (-1)^{S}P_{12}
\end{equation}
where $S$ (0 or 1) is the spin momentum of the state and $P_{12}$
the permutation operator between the two electrons. 
The $S=0$ space part of the wave function is everywhere positive, 
while the nodal surface for the 
$S=1$ state is exactly provided by the action of $\mathcal{O}$. Under this 
condition the energy can be computed by DMC without any approximation.
The singlet state supports 
a bound state and, as said above, it is thus necessary to exploit an
excited state technique. Evidently,
in order to get the required scattering information,
the boundary condition at the surface for every state must be controlled.  
As shown in Ref \cite{Shumw} the Correlation Function DMC method\cite{CepBernu} 
with our choice of $\mathcal{B}$
accomplishes automatically this requirement and it will therefore be
used in the following. Detailed descriptions of this method are
out of the scope of the present work and can be 
extensively found in the literature \cite{CepBernu,CFQMC}.
In this respect, we limit ourselves to show
the convergence for one calculation in Figure \ref{convgraph}. 
All the simulations for the triplet state of Ps-H were carried out using a time step of
0.01 hartree$^{-1}$, 2000 walkers, and a total of 100 blocks of 10000 steps each.
Simulations for the singlet states were performed employing 2000 configurations,
a time step of 0.01 hartree$^{-1}$, and a grand total of 10000 decorrelated 
Euclidean time evolutions.
Low energy phase shifts for both $S=0$ and $S=1$ systems are shown in Figure 
\ref{HPsgraph}. Scattering lengths are reported in Table \ref{HPstab},
together with three independent estimates of the same quantities very 
recently calculated \cite{Campb,Adhik,Jim}.
\begin{table}
\caption{Scattering lengths (bohr) for Ps-H scattering.}
\begin{ruledtabular}
\begin{tabular}{lcr}
 & QMC & Previous results \\
\hline
S=0 & 4.36(2) & 4.3 \cite{Jim}
, 3.49 \cite{Adhik}
, 5.20 \cite{Campb}\\
S=1 & 2.24(1) & 2.2 \cite{Jim}
, 2.46 \cite{Adhik}
, 2.45 \cite{Campb} \\
\end{tabular}
\end{ruledtabular}
\label{HPstab}
\end{table}
Whereas all of them agree in assigning the value for the triplet state,
there appears to be some controversy where the singlet state is concerned.  
Our results, which we believe to be statistically exact, are very 
close to the values proposed in Ref \cite{Jim} suggesting 
these could be safely considered as a definitive estimate.    
This fact can be also taken as a strong proof of the reliability of the method
we are employing, as well as of the full-electron Stochastic Variational Method 
approach used in Ref \cite{Jim}.

With this premise, we now address the more
debated problem of positronium scattering off helium.
Before discussing our computed quantities for this process,
it is worth noting that the experimental measurements of the threshold
value of the cross section span almost an entire order of magnitude
\cite{Skalsey,Nagash}. 
%it is worth reminding how, experimentally, 
%plenty has been said about the threshold value of the cross section, with 
%measured values
%spanning almost an entire order of magnitude \cite{Skalsey,Nagash}. 
%Theoretically the situation is not better;
The most recent theoretical estimates,
obtained by different computational schemes, do not single out
one of these as the correct one. 
%Here, we believe is worth to remind that all the previous estimates 
%generally rely on a modeling of the correlation and/or exchange effect. 
%For instance, in Ref \cite{Jim} the
%helium atom is treated as a frozen core with the suited adding of polarization
%interaction potentials, while Ref \cite{Biswas} rely on an approximated form of the 
%exchange interaction. 
%The present DMC approach should completely settle this question. 
%which should safely solve some major mathematical and computational problems. 
%The reasons of this failure are primarily due to the small value of the cross
%section which makes any approximation have a dramatic effect on the final
%computed value.  
The primary reason for this failure is the small size of the cross section
and a consequent large fractional error associated with any approximation. 
%The fractional error associated with any approximation is consequently very
%large.
To make a more direct comparison with experiments, Table \ref{HePstab} shows
the most recently calculated and measured threshold cross sections.
%instead of the corresponding phase shift.

In the present study, the system is treated
with a genuinely many-body technique and no physical approximation have been 
made prior to the numerical simulation. 
\begin{table}[b]
\caption{Scattering threshold cross section ($\pi$ bohr$^2$) for He-Ps.}
\begin{ruledtabular}
\begin{tabular}{lcr}
QMC & Experimental & Previous calculations \\
\hline
7.892(2) & 13(4) \cite{Nagash} & 13.2 \cite{Blackw} \\
         & 2.6(5) \cite{Skalsey} & 3.10 \cite{Biswas}\\
	 &                           & 10.4 \cite{Jim} \\
\end{tabular}        
\end{ruledtabular}
\label{HePstab}
\end{table}
\begin{figure}
\rotatebox{-90}{\resizebox{!}{0.45\textwidth}{\includegraphics{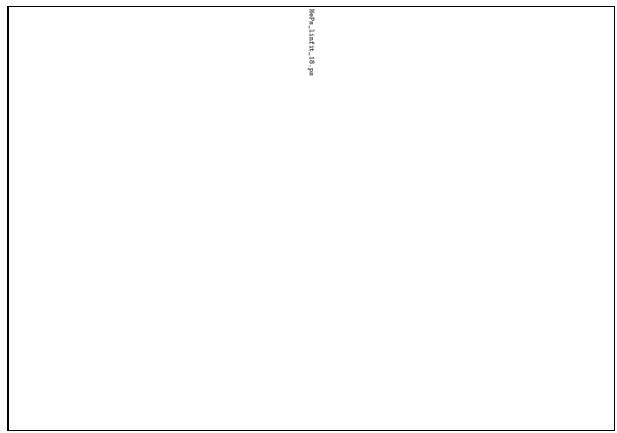}} }
\caption{Phase shift for Ps-He S-wave elastic scattering (S=1/2).
Momentum in atomic units and phase shift in radians.}
\label{distr}
\end{figure}
The absolute freedom one has in choosing the analytical form 
of the wave function in QMC methods allows us to employ the following explicitly
correlated form for $\Psi_{He}$
\begin{equation}
\Psi_{He}=\exp\left(\frac{\alpha_{1}r_{1}+\alpha_{2}r_{1}^2}{1+\alpha_{3}r_{1}}
+\frac{\beta_{1}r_{2}+\beta_{2}r_{2}^2}{1+\beta_{3}r_{2}}+\frac{\gamma_{1}r_{12}}{1+\gamma_{2}r_{12}}\right)
\end{equation}
which has a DMC energy statistically exact. Moreover, the only Young
diagram compatible with the choice of an helium atom in its ground state 
($S=0$) gives the following form for $\mathcal{O}$:
\begin{equation}
\mathcal{O}= (1+P_{12})(1-P_{13})
\end{equation} 
Simulations for this system were characterized by a time step of
0.005 hartree$^{-1}$, 4000 walkers, and a total of 130 blocks 
of 25000 steps each. The numerical results for the phase shift are shown
in Figure \ref{distr}.

The value of the computed scattering length is 
1.4046(6) bohr with a corresponding threshold cross section 
of $7.892(2) \pi a_0^2$. Comparison with numerical
estimates and experimental results curiously shows this value to lie
in a 'neutral' zone, intermediate amongst the most recently proposed values
(Table \ref{HePstab}). The nodal error, being the only approximation introduced,
deserves some comments.
General considerations \cite{Chiesa} show this bias on the phase shift 
to be always negative and proportional to $V^{-1/3}$ where $V$ is the 
sphere volume. As a result of this, our scattering length 
could be slightly lower than the exact one. More quantitatively,
one can observe that in the interaction region the employed function closely 
resembles the functional form used in bound state calculations 
on similar systems, for which the nodal error 
roughly equals $10^{-5}$ hartree \cite{maxcor}. 
In the rest of the 
simulation volume the nodes of the trial wave function are practically exact 
thanks to the validity of Eq. \ref{factform}. For such a reason we expect a 
bias on the energy of the same order of magnitude as the one in bound 
state calculations. If so, the nodal error would turn out to be smaller than
the statistical fluctuations of our energy values, therefore warranting 
the statistical exactness of our results.

Among the numerous applications directly derivable from this method, we would 
like to emphasize that the possibility to sample the exact particle 
distributions in configurational space may allow one to obtain an effective
interaction potential between Ps and a given atom or molecule. 
This potential, where all the physical effects are correctly accounted for,
could be successively used to simulate Ps in condensed phases as molecular
crystals and liquids. 
Moreover, it could also help in defining the preferential spatial
location where the Ps positron would annihilate during a "pick off" 
annihilation event, so that the interplay between the theoretical and the
experimental results may enhance the diagnostic role played by  Ps in condensed
matter science.
The DMC method is also suitable, as demonstrated in a slightly 
different context
\cite{Mella}, for the computation of $Z_{eff}$, and its formal extension to
reactive processes (inelastic are still excluded) has been known since the
seminal work of Alhassid and Koonin \cite{koonin}. 

%Before concluding a citation to the work of Drachman and Houston \cite{Drach1,Drach2,Drach3}
%is mandatory. Their estimates for both H and He, despite being almost thirty
%years old, are more accurate then many more recent results and, in particular,
%their value for the He-Ps scattering length is the only consistent with ours.

The authors are in debt to Dr. Jim Mitroy for many helpful comments and 
discussions on methodological issues and positronium physics. Financial 
support from the Universit\'a of Milano is also acknowledged.

\end{document}